\documentclass[12pt]{iopart}

\begin{document}

\title{Multiparticle production in the Glasma at NLO and plasma instabilities}

\author{Raju Venugopalan}

\address{Physics Dept., BNL, Upton, NY, USA 11973}
\ead{raju@bnl.gov}
\begin{abstract}
We discuss the relation between multi-particle production in the Glasma at next-to-leading order 
and the physics of plasma instabilities. 

\end{abstract}


\section{Introduction}
This talk is based on recent work on two topics: i) a formalism to compute multi-particle production in heavy ion collisions at next-to-leading order with Fran\c cois 
Gelis~\cite{Gelis} and ii)  3+1-D numerical simulations of Yang-Mills equations with Paul Romatschke~\cite{Paul} that investigate the role of  plasma instabilities in the  Glasma.  The word ``Glasma" describes the physics of  the strongly interacting matter from the time when particles are produced in the shattering of two Color Glass Condensates (CGCs) in a heavy ion collision to the time when the Quark Gluon Plasma (QGP) thermalizes~\cite{LappiMcLerran}. The theme here is that there exists a deep connection between 
topics i) and ii). It may provide insight into the puzzle of how the QGP is formed from the CGC  at apparently very early times of $\leq 1$ fm. 

\section{{\it Ab initio} treatment of multi-particle production in heavy ion collisions}
 
Several thousand particles are produced in a heavy ion collision at RHIC and will soon be at the LHC. Can we understand how these particles are produced and follow 
their evolution and formation of a thermalized QGP from first principles ?  Over a decade ago, heavy ion collisions were widely viewed to be an intrinsically non-perturbative 
problem not amenable to systematic computation of the properties of hot and dense matter formed. However, simple considerations have long suggested that the dynamics of 
partons at high energies are controlled by a semi-hard scale, the saturation scale $Q_s(x)$, which grows as x decreases thereby making a weak coupling description feasible: $\alpha_S(Q_s)\ll1$. 
The physics describing the dynamics of 
partons at small $x$ can be formulated as an effective field theory-the Color Glass Condensate (CGC)~\cite{IV}- which has been applied to study particle production across the 
board among high energy scattering experiments. The basic ingredients in the CGC description of the nuclear wavefunction at high energies consists of large $x$ static color 
sources $\rho(x_\perp)$ and small $x$ dynamical gauge fields. The sources are strong sources of strength $\rho\sim 1/\sqrt{\alpha_S} \gg1$. The probability of multi-particle production at high energies 
can be computed in the framework of particle production in field theories with strong external sources. Specifically, we find~\cite{Gelis}
\begin{equation}
P_n = \exp\left(-{1\over g^2}\sum_r b_r\right)\, \sum_{p=1}^n {1\over p!} \sum_{\alpha_1+\cdots \alpha_p = n} {b_{\alpha_1}\cdots b_{\alpha_p}\over g^{2p}} \, .
\label{eq:eq1}
\end{equation}
where $b_r$ denotes the sum of vacuum-to-vacuum graphs with r cuts. This formula has remarkable features which are particular to field theories with strong sources: 
a) $P_n$ is non-perturbative in $g$ even for $g\ll1$-no simple power expansion in terms of $g$ exists. b) $P_n$, for any $n$, gets contributions from cut tree vacuum graphs-this would not apply 
for field theories in the vacuum. c) Even at tree level, $P_n$ is {\it not} a Poisson distribution. This last result is non-trivial because one might naively assume that classical field 
theories would have only ``trivial" Poissonian correlations. The simple formula in Eq.~\ref{eq:eq1} is all one needs to understand many features of 
the well known AGK calculus of multi-particle production~\cite{Gelis}.

Interestingly, even though the probabilities in Eq.~\ref{eq:eq1} are non-perturbative for $g\ll1$, a systematic expansion in powers of g exists for moments of the multiplicity. 
The leading term for the first moment, the average multiplicity $\langle n \rangle= \sum_n n P_n$, is of order O($1/g^2$) in the  coupling but contains all orders in $g\rho$. The next-to-leading order 
contribution is O($g^0$) in the coupling but again all orders in $g\rho$. Another interesting feature of the moments is that even though they contain sums over r-particle Feynman graphs (corresponding 
to the $b_r$ in Eq.~\ref{eq:eq1}), they can be represented simply in terms of solutions of equations of motion with retarded boundary conditions. At leading 
order, these are solutions $A_{\rm cl.}^{a,\mu}$ to the Yang-Mills equations; the corresponding particle multiplicities are given by the relation
\begin{eqnarray}
E_p {d\langle n\rangle_{\rm LO}\over d^3 p}&=&\frac{1}{16\pi^3}\lim_{x^0,y^0\rightarrow +\infty}\int d^3 x d^3 y\,
e^{ip\cdot (x-y)}(\partial_{x^0}-iE_p)(\partial_{y^0}+iE_p)\nonumber\\
&&\times \sum_{\lambda=1,2}\varepsilon_\mu^\lambda(p){\varepsilon^\star}_\nu^\lambda (p)A_{\rm cl.}^{a,\mu}(x) A_{\rm cl.}^{a,\nu}(y)\, .
\label{eq:eq2}
\end{eqnarray}
Yang-Mills equations with {\it boost invariant} CGC initial conditions were solved numerically in Ref.~\cite{KNV} to compute the gauge fields 
at late times and the spectrum in Eq.~\ref{eq:eq2} determined. The spectrum matches the pQCD tree level expectation at large $k_\perp$ ($k_\perp > 
Q_s$). Remarkably, for small transverse momenta, the spectrum is best fit by a Bose-Einstein form. It has been suggested that such behavior is 
generic in classical glassy non-equilibrium systems~\cite{Gaglani}. Further, for any finite time, the spectrum is infrared finite. One can extract the initial energy density in a heavy ion collision from these numerical results, and one obtains
$\epsilon = 0.26\,\Lambda_S^4/ (g^2 \Lambda_S\, \tau)$.
Here $\Lambda_s$ is a scale of order $Q_s$-estimates give $\Lambda_s \sim 1.6$-$2$ GeV for RHIC energies. 
For proper times $\tau =0.3$ fm and $g=2$, this corresponds to $\epsilon \approx 20-40$ GeV/fm$^3$. In addition to providing {\it ab initio} 
estimates for the initial energy density and momentum distribution, the classical field simulations provide an estimate of the initial eccentricity 
derived from the universal properties of 
the glue fields~\cite{LappiRV}. Use of the ``universal" saturation scale as opposed to other estimates where the 
scale used is process dependent~\cite{Hiranoetal} have ramifications for the elliptic flow produced in heavy ion collisions; the elliptic flow 
generated in the Glasma will place a bound on the viscosity at RHIC.

It is important to go beyond these leading order estimates to formulate a quantitative theory of the Glasma. A next-to-leading order (NLO) computation is important to understand the renormalization and factorization issues (which are fundamental to any quantum field theory) and (as we shall discuss shortly) it is important to understand the quantum fluctuations which generate the plasma instabilities that may speed up thermalization. At NLO in 
our framework (order O($g^0$) and all orders in $g\rho$), one has two contributions to the average multiplicity~\cite{Gelis}. One corresponds to the contribution from the production of a pair of gluons while the other corresponds to the interference between the 
LO classical field (in the amplitude) and the 
one loop correction to the classical field (in the complex conjugate amplitude). Remarkably, both terms can be computed by solving
 equations of motion for small fluctuations about the classical background field with retarded boundary conditions. This means that the average multiplicity at NLO can be formulated as an {\it initial value problem} and a practical algorithm constructed to compute it. A similar algorithm has 
 been constructed and implemented to study quark pair production in the classical CGC background field~\cite{GLK}. The gluon pair production 
 NLO contribution is analogous.

 \section{The origin of instabilities in the Glasma}

In the Glasma, the classical LO {\it boost invariant} fields are purely longitudinal. The corresponding momentum distributions  (which 
only generate transverse pressure) are very unstable--
indeed, they lead to an instability analogous to the Weibel instability in electromagnetic plasmas. For a review and relevant references, see Ref.~\cite{StanMike}. The specific mechanism for a 
an expanding Glasma is that small rapidity dependent quantum fluctuations (that break the boost invariance of the underlying classical fields) 
grow exponentially and generate longitudinal pressure. Romatschke and I performed 3+1-dimensional numerical simulations of Yang-Mills equations 
for a Glasma exploding into the vacuum~\cite{Paul}. We chose an initial ``seed" spectrum of small rapidity dependent fluctuations that are ``white noise" Gaussian random fluctuations constructed to satisfy Gauss' law. We observed clearly that the maximally unstable modes of the longitudinal pressure grow as $\exp\left(C\sqrt{\Lambda_s \tau}\right)$ with $C\approx 0.425$-this form of the growth was previously predicted for Weibel instabilities in 
expanding plasmas. The distribution of unstable modes also has the ``parabolic" shape characteristic of these modes in Hard Thermal 
Loop (HTL) studies of the Weibel instability in finite T QCD. Albeit the 3+1-D solutions of the Yang-Mills equations display the same features as those of 
the HTL studies, a deeper understanding of this connection is elusive. 

The trend towards isotropization is suggested by the following observation. At early times, the unstable longitudinal momentum mode 
with the largest frequency (with amplitudes above the background mode distribution) grows linearly with proper time; this corresponds 
to times when the transverse magnetic fields generated are still growing. At the times when these begin to saturate, the maximum unstable 
mode shoots up rapidly. This is accompanied by a decrease in the transverse pressure and an increase in the longitudinal pressure. A plausible 
explanation therefore is that when the transverse magnetic fields are very large, the Lorentz force is sufficient to bend transverse momentum modes
into longitudinal modes. This behavior is seen for initial seeds with widely varying amplitudes. The energy density shows a clear deviation from free 
streaming, $\varepsilon \sim \tau^{-(1+\delta)}$ where $\delta\sim 1/16$. This value is well below the $\delta=1/3$ required in boost invariant hydrodynamics but is proof in principle that non-trivial dynamics can result from rapidity dependent quantum fluctuations. 

In an interesting recent paper, Fukushima, Gelis and McLerran~\cite{FGM} computed the first quantum corrections to the classical background 
field at $\tau=0$. The initial conditions for the subsequent classical field evolution are quite different from those in Ref.~\cite{Paul}. Simulations are 
underway to determine whether these initial conditions speed up thermalization. A full treatment of quantum fluctuations requires that we understand the 
NLO contributions to inclusive gluon production that we discussed previously-these go beyond the ``WKB" treatment in Ref.~\cite{FGM}. Some of the NLO quantum corrections can be absorbed in the evolution of the nuclear wavefunctions with energy while the rest contribute to gluon production-a 
proof of this high energy ``factorization" is desirable. A related question of great interest is the kinetic theory of the Glasma. How do the 
Glasma fields decay into particle modes and what are the corresponding kinetic equations ? Work in these directions is in progress.

This work was supported by DOE Contract No. DE-AC02-98CH10886.

\end{document}